\documentclass{caosp306}

\usepackage{graphicx}

\usepackage{natbib}
\usepackage{longtable}

\articleNo{B11}
\pubyear{2019}
\volume{49}
\volnumber{2}
\firstpage{1}
\received{November 20, 2018}
\accepted{February 1, 2019}

\def\BibTeX{{\rm B\kern-.05em{\sc i\kern-.025em b}\kern-.08em
			T\kern-.1667em\lower.7ex\hbox{E}\kern-.125emX}}

\begin{document}

\htitle{ASASSN-18fk}
\hauthor{E.\,Pavlenko, {\it et al.} }

\title{ASASSN-18fk: A new WZ Sge-type dwarf nova with multiple rebrightenings 
and a new candidate for a superhumping intermediate polar}

\author{
        E.\,Pavlenko\inst{1}  
      \and 
        K.\,Niijima\inst{2}   
      \and 
        P.\,Mason\inst{3,4}
      \and 
        N.\,Wells\inst{3,4}
      \and 
        A.\,Sosnovskij\inst{1}
      \and 
        K.\,Antonyuk\inst{1}
      \and 
        A.\,Simon\inst{5}
      \and 
        N.\,Pit\inst{1}
      \and 
        C.\,Littlefield\inst{6}
      \and 
        H.\,Itoh\inst{7}
      \and 
        S.\,Kiyota\inst{8}
      \and 
        T.\,Tordai\inst{9}
      \and 
        P.\,Dubovsky\inst{10}
      \and 
        T.\,Vanmunster\inst{11}
      \and 
        G.\,Stone\inst{12}
      \and 
        T.\,Kato\inst{2}
      \and 
        A.\,Sergeev\inst{13,14}
      \and 
        V.\,Godunova\inst{13}
      \and 
        E.\,Lyumanov\inst{1}
      \and 
        O.\,Antonyuk\inst{1}
      \and 
        A.\,Baklanov\inst{1}
      \and 
        Ju.\,Babina\inst{1}
      \and 
        K.\,Isogai\inst{2}
      \and 
        Ya.\,Romanyuk\inst{15}
      \and 
        V.\,Troianskyi\inst{16,17}
      \and 
        V.\,Kashuba\inst{17}     
       }

\institute{Crimean astrophysical observatory of RAS, Republic of Crimea,
            \email{eppavlenko@gmail.com}
         \and 
           Department of Astronomy, Kyoto University, Kyoto 606-8502, Japan
         \and 
           New Mexico State University, MSC 3DA, Las Cruces, NM, 88003, USA 
         \and
           Picture Rocks Observatory, 1025 S. Solano, Suite D, Las Cruces, NM, 88001, USA
         \and
           Astronomy and Space Physics Department, Taras Shevchenko National University of Kyiv, Volodymyrska
           str. 60, Kyiv, 01601, Ukraine
         \and
           Department of Physics, University of Notre Dame, 225 Nieuwland Science Hall, Notre Dame, Indiana
           46556, USA
         \and
          Variable Star Observers League in Japan (VSOLJ), 1001-105 Nishiterakata, Hachioji, Tokyo 192-
          0153, Japan
         \and 
          VSOLJ, 7-1 Kitahatsutomi, Kamagaya, Chiba 273-0126, Japan
         \and 
          Polaris Observatory, Hungarian Astronomical Association, Laborc utca 2/c, 1037 Budapest, Hungary
         \and 
          Vihorlat Observatory, Mierova 4, 06601 Humenne, Slovakia
         \and 
           Center for Backyard Astrophysics Belgium, Walhostraat 1A, B-3401 Landen, Belgium
         \and 
           American Association of Variable Star Observers, 49 Bay State Rd., Cambridge, MA 02138, USA
         \and 
           ICAMER Observatory of NASU, 27 Acad. Zabolotnogo str., Kyiv, 03143, Ukraine
         \and 
           Terskol Branch of the Institute of Astronomy, Russian Academy of Science, Terskol, Kabardino-Balkarian
           Republic, 361605, Russian Federation
         \and 
           Main Astronomical Observatory of the National Academy of Sciences of Ukraine, 27 Acad. Zabolotnoho str., Kyiv, 03143, Ukraine
         \and 
         Institute Astronomical Observatory, Faculty of Physics, Adam Mickiewicz University in Poznan, ul. Sloneczna 36, PL60-286 Poznan, Poland
         \and 
         Astronomical Observatory of Odessa I.I. Mechnikov National University,
         1v~Marazlievska str., Odesa, 65014, Ukraine
          }

\date{November, 20, 2018}
\maketitle

\begin{abstract}
We present the result of a multi-longitude campaign on the photometric study
of the dwarf nova ASASSN-18fk during its superoutburst in 2018.  It was
observed with 18 telescopes at 15 sites during $\sim$ 70 nights within a
three-month interval.  Observations covered the main  outburst, six
rebrightenings and 50-d decline to a near-quiescent state.  We identify
ASASSN-18fk as а WZ Sge-type dwarf nova with multiple rebrightenings and
show the evolution of the 0.06-d  superhump period over all stages of the
superoutburst.  A strong 22-min brightness modulation that superimposed on
superhumps is found during rebrightenings and decline.  Some evidence of
this modulation in a form of a sideband signal is detected during the very
onset of the outburst.  We interpret the 22-min modulation as a spin period
of the white dwarf and suggest that ASASSN-18fk is a good candidate for a
superhumping intermediate polar.
\end{abstract}

\section{Introduction}
\label{intr}
Cataclysmic variables (CVs) are close binary systems which
consist of an old (K-L) spectral type dwarf and a white
dwarf (WD). The orbital periods of most of CVs are distributed between 
$\sim$6 hours and $\sim$76-min period minimum \citep{2001cvs..book.....H}, \citep{2006MNRAS.373..484K}. There is the 2.15 -3.18 hr "period gap"  with a deficiency of CVs within it.

 The primary component, that is the WD, is
accreting matter from the secondary old-type component,
which filled its Roche Lobe and looses material via the inner
Lagrangian point see e.g. \citep{1995CAS....28.....W} for CVs in general. Depending on the primary's magnetic
field, accretion could occur through an accretion disk (non-
magnetic CVs) or accretion stream is channelled onto magnetic poles (magnetic
CVs or polars with magnetic field $B=10^{7} - 10^{8}$ G). 
Non-magnetic CVs (dwarf novae)  display outbursts. SU UMa-type dwarf novae, which occupy a region of orbital periods 76 min -- $\sim$3 hr, possess  the two types of outbursts -- the normal ones and superoutbursts that
are a result of the combination of thermal and tidal instabilities \citep{1989PASJ...41.1005O}; \citep{1996PASP..108...39O}. Typically, several normal outbursts that are shorter and slightly fainter than superoutbursts occur between two consequtive superoutbursts. The interval between superoutbursts varies between tens of days (SU UMa stars) and years -- decades (WZ Sge type stars). More on the WZ Sge-type
stars, see \citep{2015PASJ...67..108K} for the review of these stars.

During superoutbursts, there are brightness variations (superhumps) with a period of several percent longer than the orbital one. \cite{2009PASJ...61S.395K} introduced three  stages in the superhump evolution: stage A of the growing superhumps
whose period is constant and slightly longer than the period at the next stage B; stage B  with a
systematically varying period and  stage C with a shorter and almost constant period. One of the  defining characteristics of WZ Sge-type dwarf novae
are "early superhumps", the double-wave modulations during the early stage of the outburst with a period equal to the orbital one \citep{2002PASJ...54L..11K}.
Some WZ Sge-type stars may show so-called "late superhumps" -- coherent modulation during the slowly fading stage \citep{2008PASJ...60L..23K}.

The spin and orbital periods of the WD in  magnetic CVs are synchronized (with an exception of four
well-established, slightly asynchronous polars \citep{2018MNRAS.479..341P}. In the  intermediate polars (IPs)
the magnetic field of the white dwarf is $B=10^{6} - 10^{7}$ G and accretion occurs from
an accretion ring onto the magnetic poles of the highly asynchronous white dwarf. Among 1166 CVs known up to 2012  \citep{2003A&A...404..301R}, there are 38 of the  confirmed IPs with some uncertainty of WZ Sge, see  
{\small http://asd.gsfc.nasa.gov/Koji.Mukai/\-ip\-home/ip\-home.htm}; \cite{2012MNRAS.427.1004W}. The orbital periods of the known IPs are distributed between 81 and 1000 min, where the most of them have the orbital period above the period gap. Only eight IPs are placed below the period gap, which include six outbursting members \citep{2012MNRAS.427.1004W}. The only outbursting IPs with superhumps are V455 And, CC Scl and possibly WZ Sge. 

ASASSN-18fk was discovered on March 17 by the
ASASSN-team as a bright star of $12^{m}.14$. It matched 
the blue $g = 19^{m}.6$ SDSS source (VSNET-alert 21987) and according to the CRTS data showed no past outbursts. C. Littlefield reported on the 0.0570(3) d modulation (VSNET-alert 21992)
that  was preliminary identified as double-wave early superhumps of a likely new WZ Sge-type dwarf nova, ASASSN-18fk.
Here, we report results
of the multisite campaign of the ASASSN-18fk investigation during the superoutburst and its decline  up to the near-quiescent state.
%%%%%%%%%%%%%%%%%%%%%%%%%%%%%%%%%%%%%%%%%%%%%%%%%%%%%%%%%%%%%%%%%%%%%%%%%%%%%
%                       P A R A G R A P H                                    
% To generate a paragraph simply leave a blank line after the last
% sentence of the preceding paragraph as shown below.
%%%%%%%%%%%%%%%%%%%%%%%%%%%%%%%%%%%%%%%%%%%%%%%%%%%%%%%%%%%%%%%%%%%%%%%%%%%%%
\section{Observations}
The CCD-photometry of ASASSN-18fk  was done in 2018 with 18 telescopes located at 15 observatories during 70 nights (see Table 1).
All observations were obtained in unfiltered light. CCD frames
were dark subtracted and 
flat-fielded in the usual manner. Depending on a size of the telescope, time exposure, weather conditions and brightness of
the object, the accuracy of a single observation varied between $0^{m}.005$ and $0^{m}.007$. All the data were measured relative to the comparison star 148 ($\alpha_{2000}$ = 12$^{h}$08$^{m}$53.45$^{s}$,
${\delta}_{2000}$ = 19$^\circ$16'05.6") in the AAVSO designation, $V=14^{m}.813$, $B-
V=0^{m}.792$ from the sequence X23128MX (AAVSO) and expressed in 
the Heliocentric Julian Day (HJD). During the time of our observations, the brightness of ASASSN-18fk decreased from $13^{m}.3$ to $19^{m}.5$.

\newpage
\footnotesize
\begin{longtable}{lccr}
	\caption{Journal of observations.}\\
	\hline\hline
	HJD 2458000+ (start - end) & Observatory/telescope  & CCD & N\\
	\hline
	\endfirsthead
	\caption{Journal of observations (continued).}\\
	\hline
	HJD 2458000+ (start - end) & Observatory/telescope  & CCD & N\\
	\hline
	\endhead
	\hline
	\endfoot
	\hline\hline
	\endlastfoot
	
	195.587 - 195.821 & LCO/0.80m & SBIG STL-1001E & 1543 \\
	199.409 - 199.599 & DPV/1m & MII G2-1600 & 244 \\
	199.566 - 199.799 & LCO/0.80m & SBIG STL-1001E & 1078 \\
	200.287 - 200.440 & DPV/1m & MII G2-1600 & 196 \\
	200.303 - 200.408 & Trt/0.25m & ALCCD 5.2 (QHY6) & 263 \\
	200.570 - 200.682 & LCO/0.80m & SBIG STL-1001E & 520 \\
	201.423 - 201.627 & DPV/1m & MII G2-1600 & 263 \\
	201.501 - 201.572 & Trt/0.25m & ALCCD 5.2 (QHY6) & 135 \\
	202.048 - 202.322 & Ioh/0.25m & SBIG ST-9XE & 440 \\
	202.077 - 202.211 & Kis/0.25m & APOGEE F47 & 263 \\
	202.293 - 202.408 & Trt/0.25m & ALCCD 5.2 (QHY6) & 544 \\
	202.388 - 202.511 & Van/0.40m & SBIG ST-10XME & 89 \\
	202.401 - 202.610 & DPV/1m & MII G2-1600 & 254 \\
	202.935 - 202.999 & Kis/0.25m & APOGEE F47 & 117 \\
	203.000 - 203.209 & Kis/0.25m & APOGEE F47 & 480 \\
	203.003 - 203.290 & Ioh/0.25m & SBIG ST-9XE & 455 \\
	203.227 - 203.534 & CrAO/0.38m & APOGEE E47 & 436 \\
	203.976 - 203.999 & Ioh/0.25m & SBIG ST-9XE & 39 \\
	204.000 - 204.030 & Ioh/0.25m & SBIG ST-9XE & 49 \\
	204.037 - 204.240 & Kis/0.25m & APOGEE E47 & 468 \\
	204.937 - 204.999 & Ioh/0.25m & SBIG ST-9XE & 80 \\
	205.000 - 205.031 & Ioh/0.25m & SBIG ST-9XE & 51 \\
	206.024 - 206.233 & Kis/0.25m & APOGEE E47& 484 \\
	212.227 - 212.453 & CrAO/0.38m & APOGEE E47&160\\ 
	213.224 - 213.547 & CrAO/0.38m & APOGEE E47&216\\
	215.311 - 215.584 & CrAO/0.38m & APOGEE E47&193\\
	216.238 - 216.503 & CrAO/0.38m & APOGEE E47&188\\
	217.284 - 217.445 & CrAO/0.38m & APOGEE E47&114\\
	219.229 - 219.338 & CrAO/2.6m& APOGEE E47&733\\
	220.247 - 220.379 & CrAO/2.6m& APOGEE E47&724\\
	221.288 - 221.425 & CrAO/1.25m& ProLine PL23042&847\\
	222.279 - 222.377 & CrAO/1.25m& ProLine PL23042&250\\
	223.263 - 223.390 & CrAO/1.25m& ProLine PL23042&86\\
	224.270 - 224.387 & CrAO/1.25m& ProLine PL23042&293\\
	226.297 - 226.446 & CrAO/1.25m& ProLine PL23042&71\\
	228.303 - 228.310 & CrAO/1.25m& ProLine PL23042&19\\
	229.262 - 229.380 & CrAO/1.25m& ProLine PL23042&272\\
	230.264 - 230.392 & CrAO/1.25m& ProLine PL23042&89\\ 
	230.395 - 230.539 & Terskol/0.6m& SBIG STL-1001&296\\ 
	231.249 - 231.478 & Terskol/0.6m& SBIG STL-1001&195\\ 
	231.257 - 231.390 & CrAO/1.25m& ProLine PL23042&179\\ 
	232.362 - 232.484 & CrAO/1.25m& ProLine PL23042&165\\
	233.296 - 233.420 & Terskol/0.6m& SBIG STL-1001&269\\
	233.309 - 233.461 & CrAO/1.25m& ProLine PL23042&72\\
	234.290 - 234.373 & CrAO/1.25m& ProLine PL23042&35\\
	235.363 - 235.465 & CrAO/1.25m& ProLine PL23042&138\\
	236.356 - 236.378 & CrAO/1.25m& ProLine PL23042&31\\
	237.338 - 237.353 & CrAO/1.25m& ProLine PL23042&41\\
	238.288 - 238.308 & CrAO/1.25m& ProLine PL23042&45\\
	239.270 - 239.397 & CrAO/1.25m& ProLine PL23042&90\\
	239.291 - 239.401 & Lisnyky/0.35m& SBIG ST-8XMEI&88\\ 
	240.338 - 240.340 & Lisnyky/0.7m& ProLine PL4710&6\\ 
	240.346 - 240.352 & Terskol/0.6m& SBIG STL-1001&222\\ 
	241.311 - 241.314 & Lisnyky/0.7m& ProLine PL4710&5\\ 
	242.282 - 242.412 & CrAO/0.38m& APOGEE E47&63\\ 
	242.324 - 242.382 &Lisnyky/0.7m& ProLine PL4710&166\\
	243.290 - 243.428 & CrAO/0.38m& APOGEE E47&66\\
	243.307 - 243.311 &Lisnyky/0.7m &ProLine PL4710&5\\
	244.414 - 244.506 & Mayaky/0.8m&MicroLine 9000 &104\\
	245.251 - 245.525 & CrAO/0.38m& APOGEE E47&127\\
	249.315 - 249.507 & Lisnyky/0.35m& SBIG ST-8XMEI&179\\ 
	251.285 - 251.343 & CrAO/1.25m& ProLine PL23042&24\\
	252.289 - 252.359 & CrAO/1.25m& ProLine PL23042&34\\
	253.413 - 253.487 & Mayaky/0.8m&MicroLine 9000 &72\\
	254.295 - 254.400 & CrAO/1.25m& ProLine PL23042&49\\
	259.331 - 259.380 & CrAO/1.25m& ProLine PL23042&24\\
	260.278 - 260.452 & CrAO/2.6m& APOGEE E47&259\\
	261.278 - 261.310 & CrAO/2.6m& APOGEE E47&43\\
	278.288 - 278.377 & CrAO/2.6m& APOGEE E47&121\\
	279.282 - 279.359 & CrAO/2.6m& APOGEE E47&107\\
	287.641 - 287.731 & McDonald/2.1m& ProEM&1560\\         
	\hline
\end{longtable}

\small

{\bf Description of columns:}

HJD 2458000+ (start-end): beginning and end of an observational run. 

Observatory/telescope: 
LCO - C.Littlefield,
DPV - P.Dubovsky,
Trt - T.Tordai,
Ioh - H.Itoh,
Kis - S.Kiyota,
Van - T.Vanmunster;
CrAO - Crimean Astrophys. Obs.,
Lisnyky - Lisnyky Obs.,
Terskol - Terskol Obs., 
Mayaky - Mayaki Obs.,
McDonald - Picture Rocks Obs.

CCD: CCD camera type.
N:  number of observations. 

\vspace{2mm}

\section{Superoutburst and superhumps}
\begin{figure}
	\centerline{\includegraphics[width=0.65\textwidth,clip=]{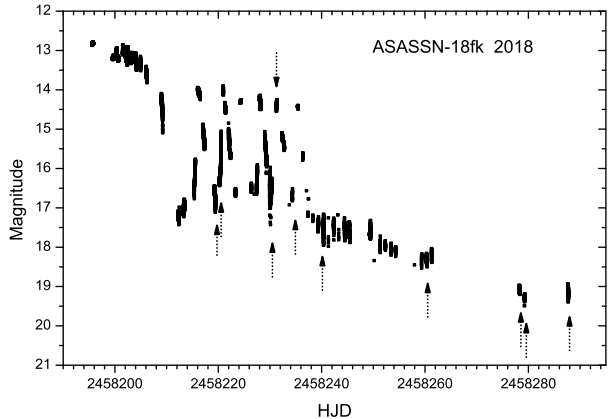}}
	\caption{The overall light curve. The dates when the 0.015-d period was detected are shown by arrows.}
	\label{LC1}
\end{figure}

The overall light curve of ASASSN-18fk during the superoutburst is presented in Fig. \ref{LC1}. It includes the main outburst, six rebrightenings and the $\sim$50-d decline to the near-quiescent state. Taking into account the ASASSN data, one could conclude that the superoutburst amplitude was about $7^{m}$ and the main outburst lasted for $\sim$ 15 d. The amplitude of rebrightenings was $\sim 3^{m}$. It seems that $\sim$ 100 d after the start of the outburst, ASASSN-18fk reached its
quiescence, or at least appeared close to it.
 \begin{figure}
	\centerline{\includegraphics[width=0.55\textwidth,clip=]{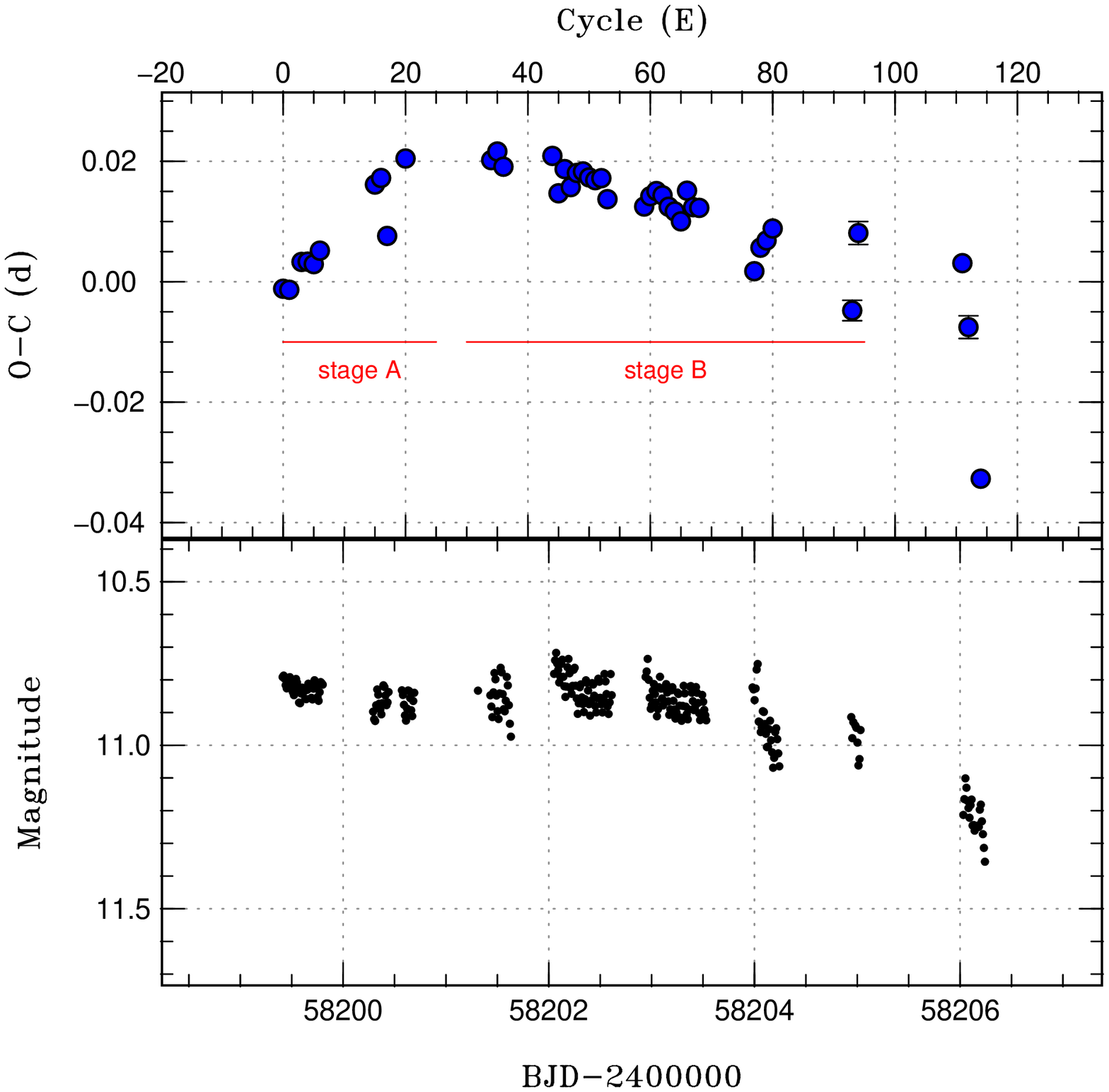}}
	\caption{Evolution of superhumps. Upper panel: O-C for the superhump maxima. Stages A and B are marked. Lower panel: a part of the main outburst.}
	\label{oca18fk1}
\end{figure}
In the course of the main outburst, rebrightenings and superoutburst decline, the superhumps have been observed. The early superhumps, having a two-humped profile, were detected only for JD 2458195 with a period of 0.057-0.060 d.
A further  gap in the observations did not allow us to estimate the duration of this stage. 
Using both the periodogram analysis \citep{1992isda.book.....P} and O-C method for the superhump maxima, we identified  a stage of growing superhumps A, with a period of 0.06075(2) d  and a stage of fully developed superhumps B, with a near-constant mean period of 0.05940(1) d. The stage A lasted for at least about 20
cycles,  while stage B -- about 60 cycles (see Fig. \ref{oca18fk1}). 

The lack of data at the stage of early superhumps did not ensure sufficient accuracy in estimating the period that could be considered as the orbital one, and, hence, in defining
the binary mass ratio, using the method proposed by \cite{2013PASJ...65..115K}.  During the stages of rebrightenings and superoutburst decline, the period of superhumps was 0.059586(7) d and 0.059521(4) d, respectively, that is 0.3$\%$ - 0.2$\%$ larger
than the period during stage B. We suggest that this period probably could be a period of late superhumps similar to what was observed in three WZ Sge-like stars GW Lib, V455 And and WZ Sge \citep{2008PASJ...60L..23K}. No orbital periodicity was detected after the main outburst termination. 

Taking into account the outburst features described, ASASSN-18fk could be defined as a WZ Sge-type dwarf nova with multiple rebrightenings -- type B outburst according to the classification given by \cite{2015PASJ...67..108K}. 

\section{Short-periodic variations}

We found that additionally to the 0.06-d superhumps the nights of the best quality displayed a short-term 22-min brightness variability at different stages of the superoutburst -- during rebrightenings and  superoutburst decline.  This modulation was not seen during the main outburst.

\begin{figure}
	\centerline{\includegraphics[width=0.75\textwidth,clip=]{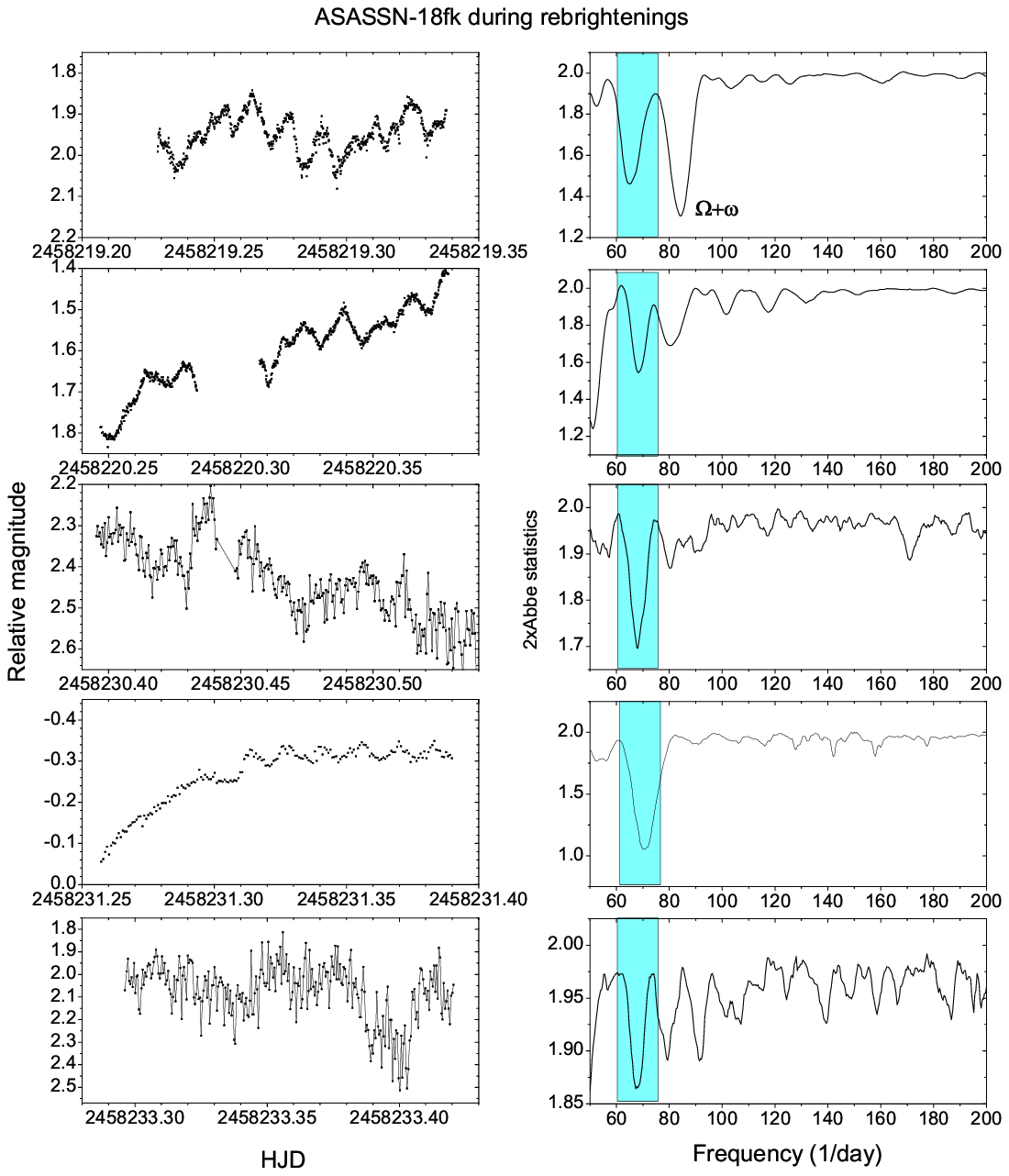}}
	\caption{Left: examples of original nightly light curves for the rebrightening stage displaying the 0.015-d period. Right: corresponding periodograms. The region around the 0.015-d period is marked by the blue strip.}
	\label{spin-R1}
\end{figure}

\begin{figure}
	\centerline{\includegraphics[width=0.75\textwidth,clip=]{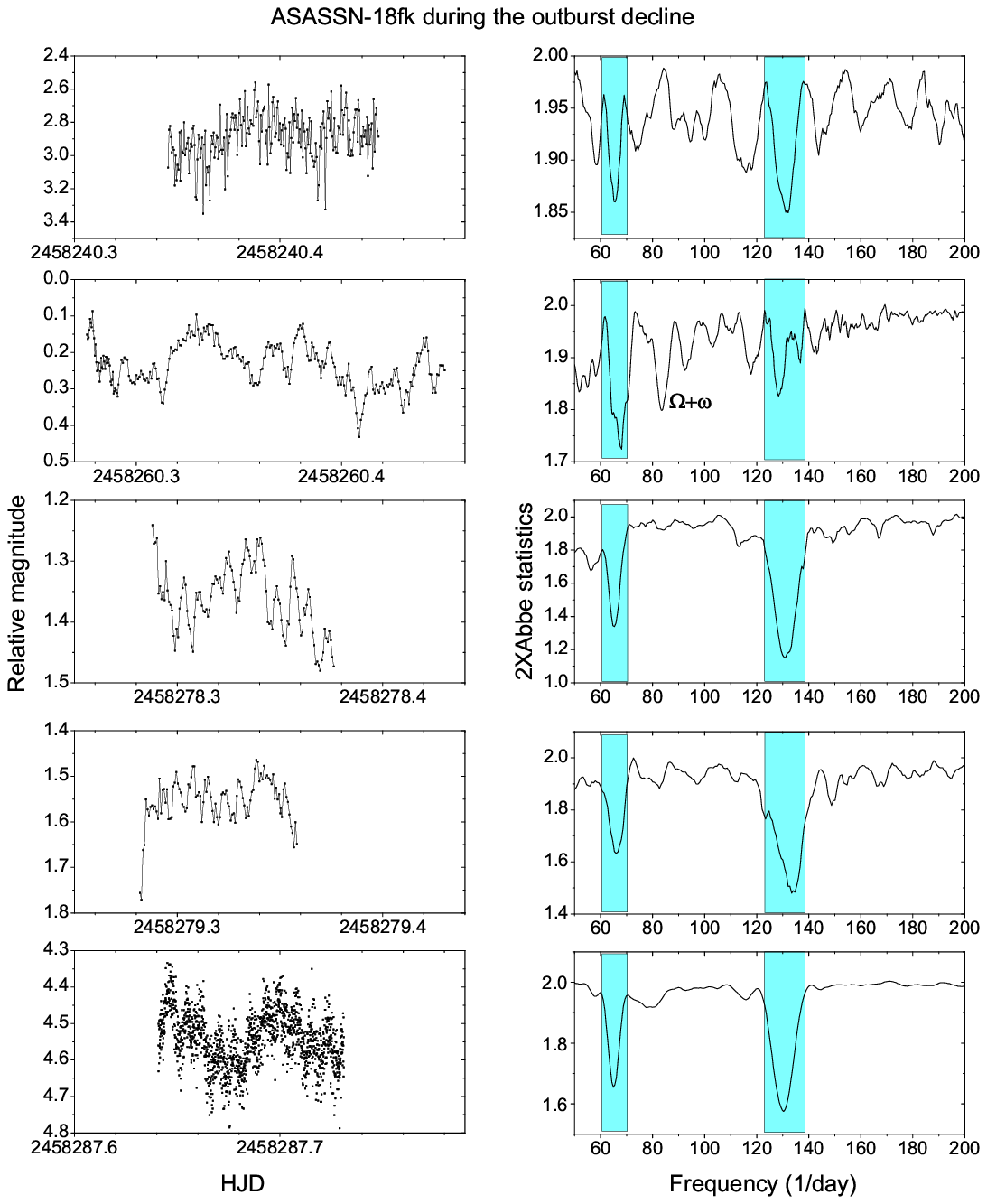}}
	\caption{Left: examples of original nightly light curves for the superoutburst decline stage displaying the 0.015-d period. Right: corresponding periodograms.The regions around the 0.015-d period and its first harmonic are marked by the blue strip.}
	\label{spin-D1}
\end{figure}

We considered a short-term periodicity for these stages  separately. We constructed the periodograms for these selected data using the ISDA package \citep{1992isda.book.....P} after removing a superhump wave and a trend corresponding to the superoutburst profile. The original light curves for these nights and  periodograms for the detrended data are shown in Fig. \ref{spin-R1}, and in Fig. \ref{spin-D1}, for the outburst decline stage. All the periodograms contain the strongest peak around 0.015 d (22 min) with an ecxeption of the first night where the 0.015-d period has the second rating. Contrary to the 0.015-d period that was detected at the rebrightening stage, these data contain a signal at the first harmonic of the 0.015-d period which became stronger as ASASSN-18fk approached the quiescence.

\begin{figure}
	\centerline{\includegraphics[width=0.5\textwidth,clip=]{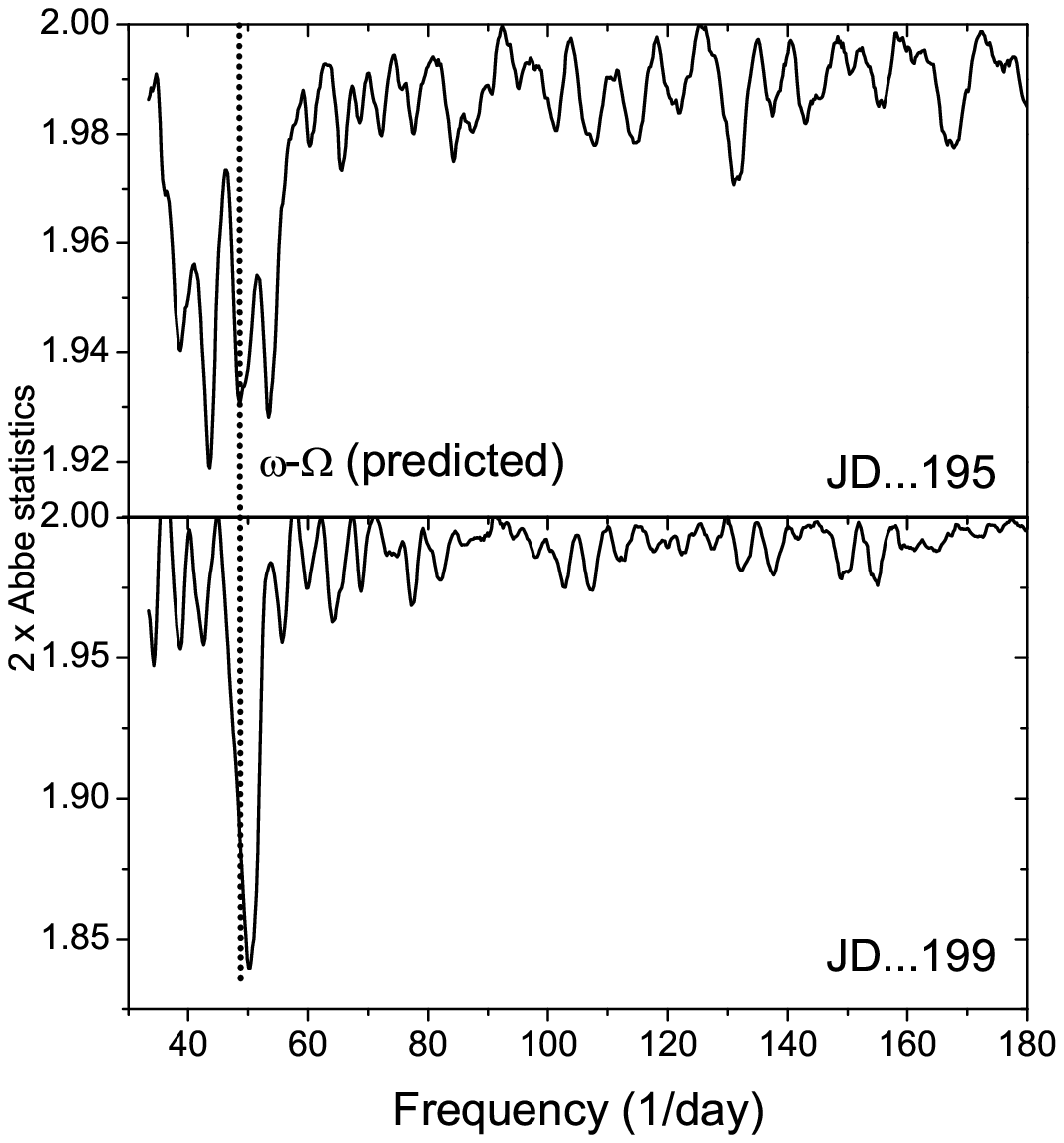}}
	\caption{Periodograms for the data during two nights of observations: JD 2458195 (above) and JD 2458199 (below). The dotted line indicates a position of the side-band    $\omega-\Omega$ frequency. The last three numbers of JD are given for the date of observations.}
	\label{beat1}
\end{figure}

We assume that these short-periodic variations may be related to the spin period of the WD. Therefore, ASASSN-18fk could be classified as an intermediate polar (IP). Another  evidence  in favor of the IP is a modulation around $\sim$50$ d^{-1}$ (for JD=2458195 and JD=2458199) and at $\sim$84$ d^{-1}$ (for JD=2458219 and JD=2458260) that coincides with beat frequencies $\omega-\Omega$ and $\omega+\Omega$ between 0.06-d and 0.015-d periods (see Fig.\ref{beat1}, Fig.\ref{spin-R1} and Fig.\ref{spin-D1}), where $\omega$ and $\Omega$ are spin and orbital frequencies, respectively. The possibility of existence of the spin-orbital sidebands of  $\omega \pm \Omega$ for IPs was predicted earlier by \cite{1986MNRAS.219..347W} and \cite{1992MNRAS.255...83W}. Note that
although a precise orbital frequency is unknown, we could roughly  admit it to be close to the 0.06-d superhump period. 

\section{The place of ASASSN-18fk among outbursting IPs}

Due to our preliminary classification, several questions arise concerning  the  period coherence, its profile, change of amplitude in the range from rebrightenings to the late decline and place in a space of orbital -- spin periods among  outbursting IPs.

\subsection{Is the 0.015-d periodicity coherent throughout the main outburst, rebrightenings and outburst decline?}  

\begin{figure}
	\centerline{\includegraphics[width=0.45\textwidth,clip=]{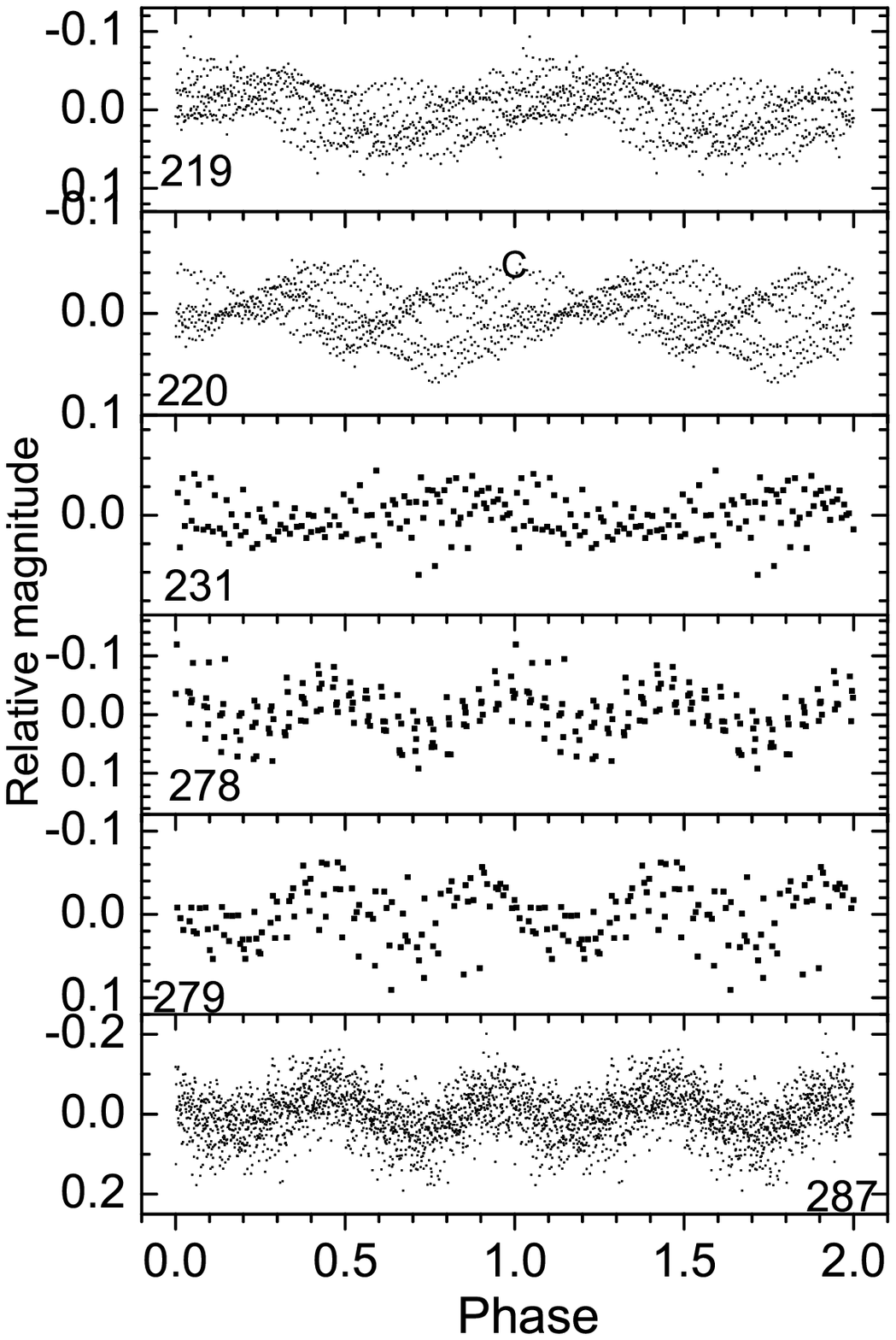}}
	\caption{The detrended data folded on the 0.0153492-d period with zero-epoch HJD=2458219.22859. The last three numbers of JD are given for the date of observations. The light curves from the top to bottom correspond to the observations between the first and second rebrightening (JD ...219), during the rise to the second rebrightening (JD ...220), during the top of the fifth rebrightening (JD ...231) and during the superoutburst decline (JD ...278, ...279 and ...287).}
	\label{coherence2}
\end{figure}

Fig. \ref{coherence2} shows the examples of the phase light curves folded on the  0.015-d period for the data between rebrightenings, at the top of the rebrightening and close to the quiescent state.

It is seen that the 0.015-d modulation was highly coherent during the outburst decline, but it was not coherent during the stage of rebrightenings.  The magnetosphere radius of the white dwarf varied on a scale of days during rebrightenings
and, as a whole, it was smaller than during the late stage of the outburst decline, when the magnetospere was rather "quiet". Obviously, it will produce both different accretion geometry and visibility of accretion regions between these two stages. It is known that another IP, DO Dra, displayed an unstable profile of the spin period and a significant shift of the peak at the spin frequency in periodograms  for some nights \citep{2002AJ....123..413S}; \citep{2008A&A...486..855A}.

\subsection{Is a difference between spin profiles at high and low brightness stages common for the IPs?}

As shown in Section 4, ASASSN-18fk displayed a one-humped spin profile during rebrightenings and a two-humped one during the outburst decline. Most of the known outbursting IPs  exhibit the same behavior. One explanation of this phenomenon was given for XY Ari by \cite{1997MNRAS.292..397H}: humps on the spin light curve are associated with two accretion regions on a white dwarf. In quiescence, the magnetosphere radius is rather large and two zones are visible. In outburst, the magnetosphere is reduced by a higher accretion rate and the accretion disk obscures one of the poles. However, as noted by \cite{2001cvs..book.....H}, "Currently, we do not have enough observational clues to determine why some systems show double-peaked pulses while others are single peaked". 

\subsection{What is an amplitude of the spin pulse at different stages of the superoutburst?}

\begin{figure}
 	\centerline{\includegraphics[width=0.6\textwidth,clip=]{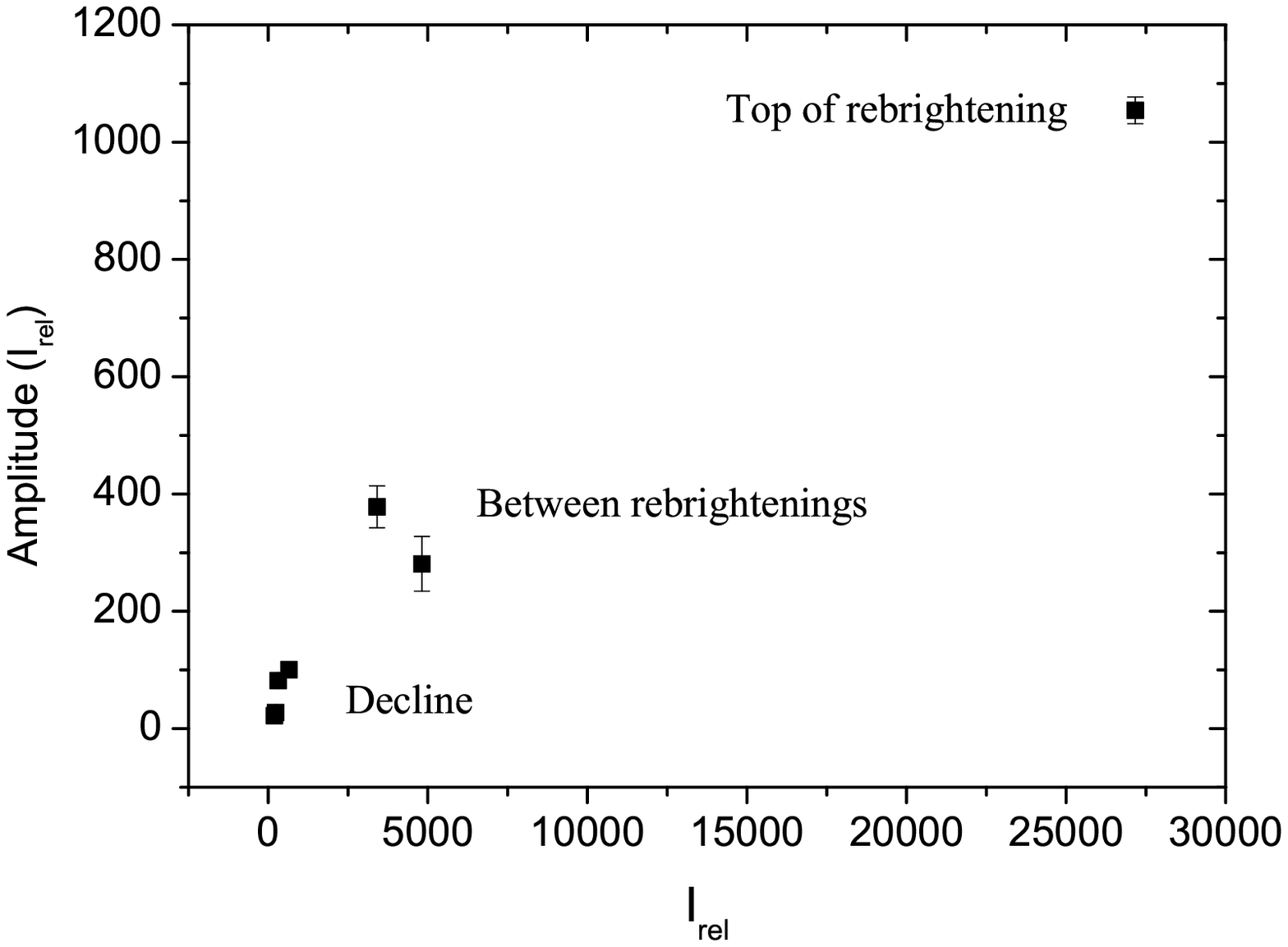}}
 	\caption{Dependence of the amplitude of the 0.015-d modulation expressed in the relative intencity ($I_{rel}$) on the mean relative intensity from the decline to the top of rebrightening ($I_{rel}=10^{10}\times10^{-0.4m}$). }
 	\label{Amp-I1}
 \end{figure}

 The mean amplitude of the 0.015-d period was about $0^{m}.04$ at the stage of rebrightenings and $0^{m}.09$ during the gradual decline stage. Converting magnitudes into relative intensities, we found a dependence of the amplitude of the mean spin pulse  of ASASSN-18fk on the mean brightness (see Fig. \ref{Amp-I1}). At the top of the rebrightening, the spin amplitude  was three times larger than those between rebrightenings and $\sim$50 times larger than those at the end of the outburst decline.  The tendency of
a growing amplitude with brightness is not  common for all outbursting IPs. Thus, there is no significant difference in the spin amplitudes  for outbursts and quiescence for DO Dra, but there is a similar dependence  for XY Ari \citep{1997MNRAS.292..397H}. 
 
\subsection{What is a position of ASASSN-18fk in the $P_{orb}$ - $P_{spin}$ diagram?}

\begin{figure}
	\centerline{\includegraphics[width=0.6\textwidth,clip=]{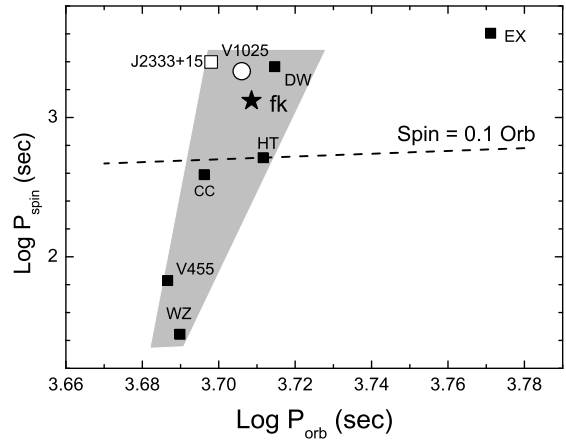}}
	\caption{Dependence between spin and orbital periods for the IPs below the period gap. Abbreviations mean: EX = EX Hya; J2333+15 = SDSS J2333+15; V1025 = V1025 Cen; DW = DW Cnc; HT = HT Cam, V455 = V455 And; CC =  CC Scl; fk = ASASSN-18fk. SDSS J2333+15 with no information on possible outburst and V1025 Cen are marked by the open square and circle, respectively. The rest are  outbursting IPs. ASASSN-18fk is marked by
a star symbol. The dashed line corresponds to the $P_{spin} = 0.1P_{orb}$ relation. Gray area indicates location of  IPs below the period gap (with an exception of EX Hya).} 
	\label{spin-orb1}
\end{figure}

We used the available data \citep{2012MNRAS.427.1004W}
 and our estimate for ASASSN-18fk to plot a dependence of the spin period on the orbital one for all  IPs below the period gap. Taking into account that the expected orbital period for ASASSN-18fk is slightly lower than the superhump period, we could use the mean superhump period value as an estimate of the orbital one. In this case the spin period is $\sim$3.8 times
smaller than the orbital one, that is close to the values of the known  IPs below the period gap \citep{2012MNRAS.427.1004W}. The position of ASASSn-18fk among these IPs is shown in Fig. \ref{spin-orb1}. While IPs above the period gap are on or below the line of $P_{spin} = 0.1P_{orb}$
\citep{2001cvs..book.....H}, IPs with the shortest orbital period (excepting EX Hya) are concentrated within a strip on either side of this line. ASASSN-18fk is situated well within this strip.  
\section{Conclusion}

Our main findings for ASASSN-18fk are as follows: 

1) ASASSN-18fk is a WZ Sge-type dwarf nova with six rebrightenings and a superhump period of 0.06-d; 2 )the one-humped 22-min signal was detected during the rebrightening stage and the two-humped one -- during the outburst decline,  which is assumed to be  the spin period of the white dwarf; 3) spin-orbital sideband periods during the main outburst.

Thus ASASSN-18fk is a first IP among the rare subclass of the superhumping IPs for which evidence of the spin period
(the direct spin signal or the spin-orbital sideband signal) was  tracked through the main outburst, rebrightenings and  near-quiescent state.

\bibliography{pav_caosp306}

\end{document}